\documentclass[10pt,twocolumn,superscriptaddress,aps,pra]{revtex4-2}

\usepackage{amsmath}
\usepackage{amssymb}  
\usepackage{bbold}    
\usepackage{braket}
\usepackage{graphicx}
\usepackage{subcaption}
\usepackage{fix-cm}
\usepackage{pgfplots} 
\pgfplotsset{compat=1.5} 
\usepackage{color}
\usepackage{float}
\usepackage{ulem}
\usepackage[colorlinks=true,linkcolor=blue,citecolor=blue,urlcolor=blue]{hyperref}  

\newcommand{\e}{\mathrm{e}}
\newcommand{\psir}{\Psi_\mathrm{R}}
\newcommand{\psig}{\Psi_\mathrm{G}}
\newcommand{\rfac}{x_\mathrm{f}}

\newcommand{\mfl}[1]{\textcolor{black}{#1}} 
\newcommand{\daniel}[1]{\textcolor{black}{#1}}
\newcommand{\brady}[1]{\textcolor{black}{#1}}
\newcommand{\bradytwo}[1]{\textcolor{black}{#1}} 
\newcommand{\eva}[1]{\textcolor{black}{#1}} 
\newcommand{\tom}[1]{\textcolor{black}{#1}} 

\begin{document}

\begin{abstract}
    Rydberg atoms allow for the experimental study of open many-body systems and nonequilibrium phenomena. High dephasing rates are a generic feature of these systems, and therefore they can  {often} be described by rate equations,  {i.e.} in the classical limit.
    In this work, we 
    analyze  one potential origin of the decoherence in Rydberg atoms: dipole-force induced dephasing.  As the wave function of the Rydberg (spin-up) state  is repelled in the presence of another nearby Rydberg atom, while the ground  (spin-down) state diffuses in place, the Franck-Condon overlap between the two spin components quickly decays causing a decoherence of the spin transition. With an analytic approach we  obtain a simple expression for the dephasing rate of the Rydberg state depending on atomic and laser parameters, which agrees with numerical findings.
\end{abstract}

\title{ Dephasing in Rydberg Facilitation Due to State-Dependent Dipole Forces
}

\author{Tom Schlegel$^*$, Evangelia Konstantinidou$^*$, Michael Fleischhauer, and Daniel Brady}
\affiliation{Department of Physics and Research Center OPTIMAS, RPTU Kaiserslautern, D-67663 Kaiserslautern, Germany}

\date{\today}

\maketitle

\section{Introduction}

Many-body systems of Rydberg atoms have proven  to be an incredibly useful and versatile platform to study interacting spin systems, both in the quantum and classical regime, due to their strong interactions and long lifetimes \cite{gallagher1994rydberg}. Through recent  advances in experimental control of neutral atoms, e.g. with tweezer arrays \cite{barredo2016atom, endres2016atom, barredo2018synthetic}, Rydberg systems offer a powerful approach to investigate many-body lattice models \cite{weimer2010rydberg, vsibalic2018rydberg}. This includes simulations of the quantum spin Ising model \cite{labuhn2016tunable, bernien2017probing, schauss2015crystallization, lienhard2018observing, guardado2018probing}, topological transport properties \cite{barredo2015coherent, de2019observation}, and nonequilibrium phase transitions \cite{helmrich2020signatures, wintermantel2021epidemic, brady2024griffiths, ohler2025nonequilibrium} to name a few.

Laser driven Rydberg gases often feature strong dephasing. While this is a well known feature of these systems \cite{saffman2008rabioscillations,pfau2009dephasing,lesanovsky2013microtraps}, a comprehensive understanding along with a quantitative description is largely missing.
In the present paper, we identify one mechanism responsible for such a dephasing, which is of particular relevance for Rydberg facilitation.
When regarded in the anti-blockade (facilitation) regime, excitations of Rydberg atoms can only occur in the presence of an already excited Rydberg atom \cite{ates2007antiblockade}. In tweezer arrays, this allows for the study of many-body dynamics under localization \cite{marcuzzi2017facilitation} and kinetic constraints \cite{magoni2021emergent, magoni2024coherent}. When regarding the dynamics in a gas, strong dephasing rates emerge. Consequently, the dynamics become effectively classical and very large systems can be described to great accuracy by diagonal density matrix elements, leading to classical rate equations \cite{ates2006strong, lesanovsky2013kinetic, marcuzzi2014effective, levi2016quantum}. The incoherent regime is especially well suited for the study of open systems and nonequilibrium phase transitions \cite{helmrich2020signatures, brady2024anomalous, ohler2025nonequilibrium}. In particular, in this regime experiments can be compared to large-scale numerical simulations.

 In studies of lattice spin models with Rydberg atoms, ground-state atoms are initially trapped in an optical lattice or in tweezer potentials, which are subsequently switched off during the interaction. Therefore, we consider the case of an initially localized ground state atom here. In the regime of Rydberg facilitation differential dipole forces acting on the excited and ground states cause a rapid dephasing of the transition, which we will analyze in the following.

\begin{figure}[H]
    \centering
    \includegraphics[width=\columnwidth]{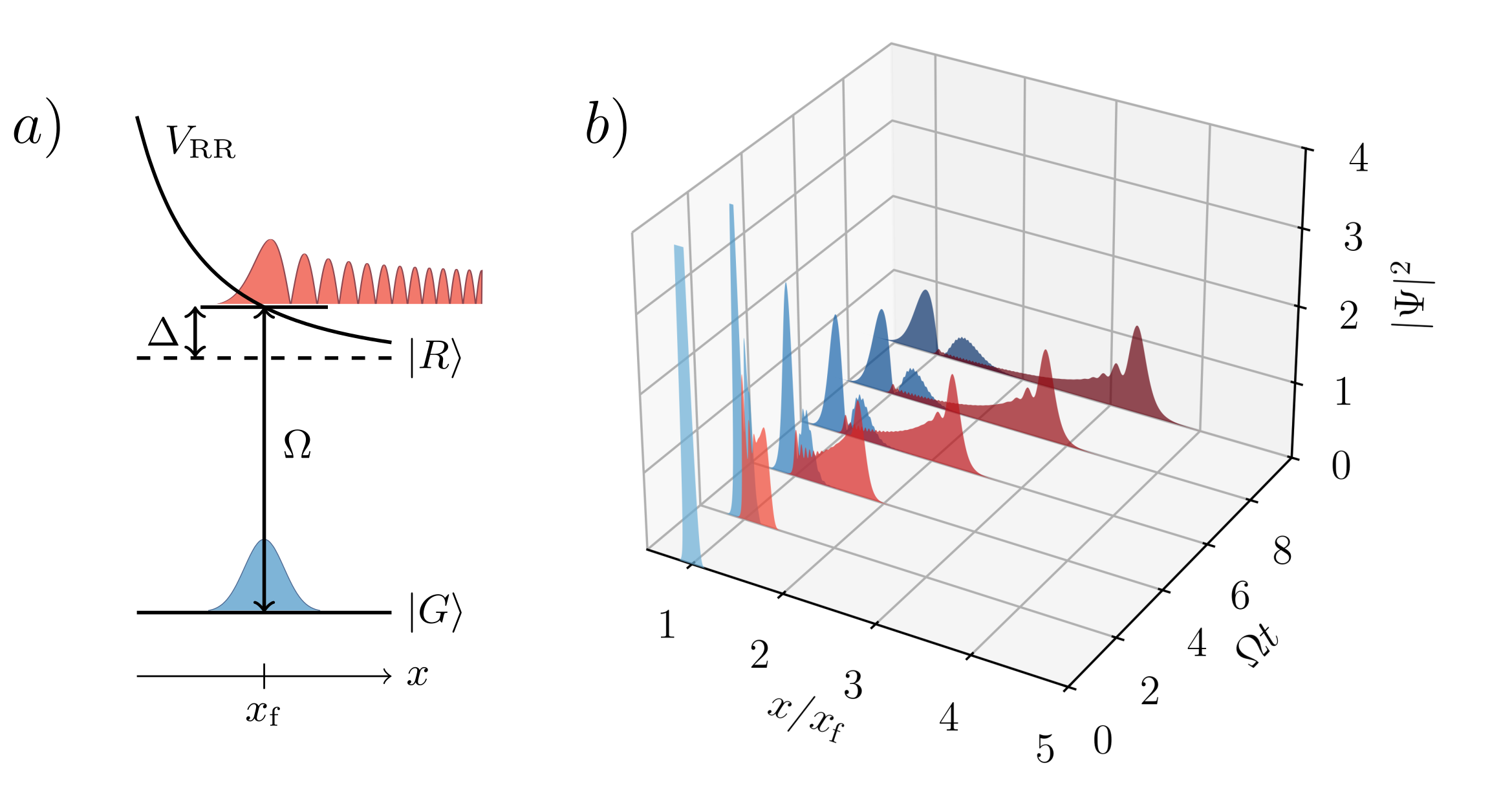}
    \caption{(a) Internal atomic structure of an atom in an external Rydberg potential $V_\mathrm{RR}$. The  {initially spatially localized} ground state $\ket{G}$ is laser-coupled with a high-lying Rydberg state $\ket{R}$ with Rabi frequency $\Omega$ and detuning from resonance $\Delta$. The Rydberg potential repels the excited part of the atom. (b) Spatially and temporally resolved dynamics of the ground (blue) and Rydberg state (red) wave packets for \brady{$\Delta / \Omega = 30$ and $\xi / \Omega = 0.01\cdot10^{-3}$.}}
    \label{fig:introduction}
\end{figure}

In the facilitation regime, the laser coupling between the internal ground $\ket{G}$ and Rydberg $\ket{R}$ states of the atom, with Rabi frequency $\Omega$, is off-resonant, with detuning from resonance ${\Delta \gg \Omega}$ (cf. Fig.~\ref{fig:introduction}a). In the absence of further couplings, the  large detuning strongly suppresses the excitation of the Rydberg state. However, in the presence of another Rydberg atom, the energy  of the excited state is shifted and becomes resonant at the interatomic distance $\rfac$, i.e. ${V_\mathrm{RR}(\rfac) - \Delta = 0}$. As a result of the repulsive van-der-Waals potential, a strong dipole force acts on the excited state. Consequently, there is a fast decay of the Franck-Condon overlap between the ground and Rydberg state wavefunctions of the atom (cf.~Fig.~\ref{fig:introduction}b).  As we quantitatively analyze in the following, this leads to an effective dephasing of the internal-state dynamics.
\mfl{The motion-induced dephasing will in general not follow a simple exponential law and the dephasing rate will be time dependent. However, for the excitation dynamics only short times after excitation are relevant and we will focus on this regime. }

\section{System}

To investigate the \mfl{short-time} dephasing dynamics, we consider a two-level atom coupled by a laser between its ground and a high-lying Rydberg state. The coupling laser has \brady{the} Rabi frequency $\Omega$ and detuning from resonance $\Delta$. Additionally, we consider the atom to be in the dipolar potential of an external Rydberg atom, located at position ${\mathbf{x}=0}$. This dipolar potential typically takes the form \daniel{${V_\mathrm{RR} = \frac{c_\nu}{|\mathbf{x}|^\nu}}$}, where $\nu$ is an integer and $c_\nu$ is the potential coefficient \cite{gallagher1994rydberg}. If both atoms are coupled to the same Rydberg state, e.g. $\ket{nS}$, then the interaction potential is of a van-der-Waals (vdW) type \cite{vsibalic2018rydberg} and ${\nu = 6}$.  In order to account for the interplay between dipole forces acting on the spatial degree of freedom of the atom in the Rydberg state, we treat the atom in second quantization with two Schr\"odinger fields $\hat \Psi_G(x)$ and $\hat \Psi_R(x)$ describing the ground and Rydberg states, respectively. Since the vdW potential only depends on the euclidean distance between the atoms, we can restrict the problem to one spatial dimension, i.e. the radial distance $x$. For the relevant time scales, diffusion in the orthogonal directions is negligible. 

The total potential $U_\nu(x)$ acting on the Rydberg state is the sum of the dipole potential $V_\mathrm{RR}$ and the detuning $\Delta$, i.e. 
\begin{equation}
    U_\nu(x) = \frac{c_\nu}{x^\nu} - \Delta.
\end{equation}
At the \textit{facilitation distance} ${\rfac \equiv \sqrt[\nu]{\frac{c_\nu}{\Delta}}}$ the interaction potential cancels the detuning and a resonant excitation of the state~$\ket{R}$ becomes possible, i.e. ${U_\nu (\rfac) = 0}$ (cf. Fig.~\ref{fig:introduction}a).

 In the following, we  consider an initial state $\vert \psi_0\rangle$ of the atom as a wave packet in the ground state with width $\sigma$ and initial position at the facilitation distance, i.e. ${x(t=0) = \rfac}$.  
%
\begin{subequations}
    \begin{align}
        \psir(x, t=0) &= \langle 0\vert \hat \Psi_R(x) \vert \psi_0\rangle = 0.
        \\
        \psig(x, t=0) &= \langle 0\vert \hat \Psi_G(x) \vert \psi_0\rangle = \frac{1}{(\pi \sigma^2)^{\frac{1}{4}}}\,\mathrm{e}^{-\frac{(x - \rfac)^2}{2 \sigma^2}}
    \end{align}
\end{subequations} 
Under the time-dependent Schrödinger equation, the equations of motion for the ground and Rydberg state wave packets are given as
\begin{equation}
    \label{eq:CoupledODE}
    i\partial_t
    \begin{pmatrix}
        \psir \\ \psig
    \end{pmatrix}
    =
    \begin{pmatrix}
        -\frac{\partial^2_x}{2m} + U_\nu(x) & \Omega \\
        \Omega    & -\frac{\partial^2_x}{2m} \\ 
    \end{pmatrix}
    \begin{pmatrix}
        \psir \\ \psig
    \end{pmatrix},
\end{equation}
where we set ${\hbar = 1}$.

\section{Analytical Approach}

\tom{In the following, we derive an analytic expression for \bradytwo{the} \mfl{short-time evolution of} the coherence $\rho_\mathrm{RG}(t)$ from which we extract a dephasing rate $\gamma_\perp$.} To this extent, we linearize the potential \tom{$U_\nu(x)$} at the facilitation distance $\rfac$ and \brady{define} $y\equiv(x-\rfac)/\rfac$. Applying this to the time-dependent Schrödinger equation~\eqref{eq:CoupledODE}, we obtain
\begin{equation}
    \label{eq:CoupledODE_2}
    i\partial_t
    \begin{pmatrix}
        \psir \\ \psig
    \end{pmatrix}
    =
    \begin{pmatrix}
        -\xi\partial^2_y -\nu\Delta y & \Omega \\
        \Omega    & -\xi\partial^2_y \\ 
    \end{pmatrix}
    \begin{pmatrix}
        \psir \\ \psig
    \end{pmatrix},
\end{equation}
where we defined $\xi\equiv1/2m\rfac^2$. This allows us to treat the system in a perturbative approach with the linearized potential $-\nu\Delta y$ as the perturbation under the assumptions $\sigma\ll \rfac$ and ${|y|\ll \big|\frac{\Omega}{\nu \Delta } \big|}$. The latter constraint also restricts this approach to be valid only for short times as a result of diffusion and dipolar repulsion. \tom{In addition, this implies an upper limit to the ratio ${\Delta / \Omega}$.} 

The unperturbed system\daniel{, i.e. $-\nu\Delta y = 0$,} can be solved exactly in $k$-space. It obeys Rabi-oscillations between the ground and Rydberg states, as expected, with the solution given by
\begin{equation}
    \label{eq:0th-order}
    \widetilde{\Psi}^{(0)}_{\mathrm{R/G}}(k,t) = \widetilde{\Psi}_\mathrm{G}(k,0)\,\mathrm{e}^{-i\xi k^2t}
    \left\{\begin{matrix}
    -i\sin\Omega t\ \\ \cos\Omega t
    \end{matrix}\right.,
\end{equation}
where $\widetilde{\Psi}^{(0)}_{\mathrm{R/G}}(k,t)$ correspond to the unperturbed solutions. In first order perturbation theory, the time-dependent Schrödinger equation in $k$-space reads 
%
%
%
\begin{equation}
    \begin{split}
        \label{eq:1st-order-ODE}
        i\partial_t
        \begin{pmatrix}
            \widetilde{\Psi}^{(1)}_\mathrm{R} \\ \widetilde{\Psi}^{(1)}_\mathrm{G}
        \end{pmatrix}
        =
        &\begin{pmatrix}
            \xi k^2 & \Omega \\
            \Omega & \xi k^2 \\ 
        \end{pmatrix}
        \begin{pmatrix}
            \widetilde{\Psi}^{(1)}_\mathrm{R} \\ \widetilde{\Psi}^{(1)}_\mathrm{G}
        \end{pmatrix}
        \\
        &+ i\nu\Delta k \left(\frac{\sigma^2}{\rfac^2}+{i2\xi t}\right)
        \begin{pmatrix}
            \widetilde{\Psi}^{(0)}_\mathrm{R} \\ 0
        \end{pmatrix}
        .
    \end{split}
\end{equation}
We can solve eq.~\eqref{eq:1st-order-ODE} exactly using variation of constants and receive
\begin{subequations}
    \begin{align}
        \notag
        &\widetilde{\Psi}^{(1)}_\mathrm{R}(k,t) = \widetilde{\Psi}_\mathrm{G}(k,0)\,\mathrm{e}^{-i\xi k^2t}\Bigl[-i\sin\Omega t +\tfrac{\nu\Delta k}{2\Omega}\\
        &\cdot\left(\tfrac{\xi}{\Omega}\sin\Omega t-\xi t\cos\Omega t-i\Omega(\tfrac{\sigma^2}{\rfac^2}t+i\xi t^2)\sin\Omega t\right)\Bigr] \label{eq:1st-ryd}
        \\
        \notag
        \\
        \notag
        &\widetilde{\Psi}^{(1)}_\mathrm{G}(k,t) = \widetilde{\Psi}_\mathrm{G}(k,0)\,\mathrm{e}^{-i\xi k^2t}\Bigl[\cos\Omega t -\tfrac{\nu\Delta k}{2\Omega}\\
        &\cdot\left(\tfrac{\sigma^2}{\rfac^2}\sin\Omega t+i\xi t\sin\Omega t-\Omega(\tfrac{\sigma^2}{\rfac^2}t+i\xi t^2)\cos\Omega t\right)\Bigr]. \label{eq:1st-gr}
    \end{align}
\end{subequations}
\tom{Inserting the inverse Fourier transform $\Psi_\mathrm{R/G}(y,t) = \frac{1}{\sqrt{2\pi}}\int\mathrm{d}k\,\widetilde{\Psi}_\mathrm{R/G}(k,t)\,\mathrm{e}^{iky}$, the coherence \mfl{then reads}
\begin{equation}
    \label{eq:RhoRG}
    \rho_\mathrm{RG}(t) = \rfac\int \mathrm{d}k\, \widetilde{\Psi}^*_\mathrm{R}(k,t) \widetilde{\Psi}_\mathrm{G}(k,t).
\end{equation}
This finally yields an analytical expression for the coherence $\rho_\mathrm{RG}(t)$ and up to first order in $t$ it is}
\begin{subequations}
    \label{eq:Rho_analytic}
    \begin{align}
        \label{eq:RhoReal}
        \mathrm{Re}[\rho_\mathrm{RG}] = &-\frac{\nu^2 \Delta^2}{8\Omega}\left(\frac{\xi}{\Omega^2}\sin^2\Omega t-\frac{2\xi}{\Omega}t\sin\Omega t\cos\Omega t\right) \notag
        \\
        &+\mathcal{O}(t^2)
    \\
        \label{eq:RhoImag}
        \mathrm{Im}[\rho_\mathrm{RG}] = &\,\sin\Omega t\cos\Omega t-\frac{\nu^2 \Delta^2}{8\Omega}\left(\frac{\sigma^2}{\rfac^2}+\frac{\xi^2 \rfac^2}{\Omega^2\sigma^2}\right)t\sin^2\Omega t \notag
        \\
        &+\mathcal{O}(t^2).
    \end{align}
\end{subequations}
From eq.~\eqref{eq:RhoReal}, we recognize that ${\mathrm{Re}[\rho_\mathrm{RG}](t_n) = -\frac{\nu^2\Delta^2}{16m\rfac^2\Omega^3}(\frac{1}{2}-\Omega t_n)}$, where ${\Omega t_n = \frac{\pi}{4} + n\pi}$ and $n$ being an integer. Here, the real part of ${\rho_\mathrm{RG}}$ only yields a \mfl{negligible} contribution  compared to the imaginary part at \mfl{short} times $t_n$ (cf.~Fig.~\ref{fig:coherence-2}). This allows us to approximate ${|\rho_\mathrm{RG}|(t_n) \approx |\mathrm{Im}[\rho_\mathrm{RG}]|(t_n)}$, since $\mathrm{Re}[\rho_\mathrm{RG}]^2 \ll \mathrm{Im}[\rho_\mathrm{RG}]^2$. Consequently, we extract the dephasing rate simply from the imaginary part. Evaluating eq.~\eqref{eq:RhoImag} at the times $t_n$ yields
\begin{equation}
    \label{eq:RhoRG=Im}
    |\rho_\mathrm{RG}|(t_n)\approx\frac{1}{2}\left(1-\gamma_\perp t_n\right)+\mathcal{O}(t_n^2)
\end{equation}
with the dephasing rate
\brady{
\begin{equation}
    \label{eq:DephsingRate}
    \frac{\gamma_\perp}{\Omega}=\frac{\nu^2}{8} \left(\frac{\Delta}{\Omega
    }\right)^2\left(\frac{\sigma}{\rfac}\right)^2\left[1+ 
    \left(\frac{\rfac}{\sigma}\right)^4 \left(\frac{\xi}{\Omega}\right)^2
    \right].
\end{equation}
}
Eq.~\eqref{eq:DephsingRate} is the main result of our work. For Rydberg facilitation $\vert \Delta\vert \gg \Omega$, so the first term is large. Note, however, that we assumed $\sigma \ll \rfac$, and thus the second factor compensates the first. \tom{The impact of the atom mass is described by $\xi$ as its reciprocal value, i.e.} $\rfac^4 \mfl{\xi^2 = 1/\tom{(2m)^2}}$.
\mfl{When $m\to\infty$ the second term in the bracket vanishes. Note, however, that
for a given trapping (tweezer) potential  also $\sigma \to 0$ as $m\to\infty$.}

\section{Numerical Benchmark}

\tom{In order to \mfl{benchmark our analytic results on} the effect of the differential dipole forces on the coherences between the internal states of the atom, we perform\mfl{ed} numerical simulations.}
To this extent, we solve eq.~\eqref{eq:CoupledODE} numerically for a time interval $[0, \:t]$, discretize time in steps $\delta t$, and use a split-step Fourier algorithm \cite{hardin1973applications} to compute the unitary time evolution. The time evolution operator ${\e^{-i\hat H \delta t}}$ for the time step $\delta t$ is split according to a second-order Trotter-Suzuki decomposition \cite{hatano2005finding} as
%
 $   \e^{-i (\hat{T}+\hat{V})\delta t} = \e^{-i\hat{T}\frac{\delta t}{2}}\e^{-i\hat{V}\delta t}\e^{-i\hat{T}\frac{\delta t}{2}} + \mathcal{O}(\delta t^3) $,
%
where 
\begin{equation}
   \hat{T} = \frac{\hat{p}^2}{2m}\cdot \mathbb{1} \qquad  \hat{V} = \begin{pmatrix} U_\nu(x) & \Omega \\ \Omega & 0 \end{pmatrix}.
\end{equation}
Here, $\hat T$ corresponds to the kinetic and $\hat V$ to the potential components of the coupled partial differential equations (PDEs). In the simulation, the kinetic time evolutions are calculated in $k$-space by using a Fast-Fourier-Transformation (FFT) of the wave function. The potential term is evaluated in real space, following another FFT. However, since $\hat V$ is not diagonal, we express the time evolution operator ${\e^{-i\hat V \delta t}}$ in terms of Pauli matrices using $\e^{i\,a \hat n \cdot \hat{\vec\sigma}} = \mathbb{1} \cos a + i \,\hat n \cdot \hat{\vec\sigma} \sin a$, for a real valued $a$ and with ${|\hat n| = 1}$. Applying this onto the time evolution operator of the potential term, i.e. ${\e^{-i\hat V \delta t}}$, we receive the real space evolution \mfl{in a time step} as
\begin{subequations}
    \begin{align}
        \notag
        \psir(t+\delta t) = \ &\e^{-i\varphi}\left( \cos(\omega \delta t) -i \sin(\omega \delta t) \frac{U_\nu(x)}{2\omega} \right)\psir(t)\\
        &-i\mathrm{e}^{-i\varphi} \sin(\omega \delta t) \tfrac{\Omega}{\omega} \psig(t)\\
        \notag
        \\
        \notag
        \psig(t+\delta t) = \ &\e^{-i\varphi}\left( \cos(\omega \delta t) +i \sin(\omega \delta t) \frac{U_\nu(x)}{2\omega} \right)\psig(t)\\
        &-i \e^{-\mathrm{i}\varphi} \sin(\omega \delta t) \tfrac{\Omega}{\omega} \psir(t),
    \end{align}
\end{subequations}
\daniel{with ${\varphi = \frac{U_\nu(x)}{2}\delta t}$ and ${\omega = \sqrt{\frac{1}{4} U_\nu^2(x) + \Omega^2}}$.}


\begin{figure}[H]
    \centering
    \includegraphics[width=\columnwidth]{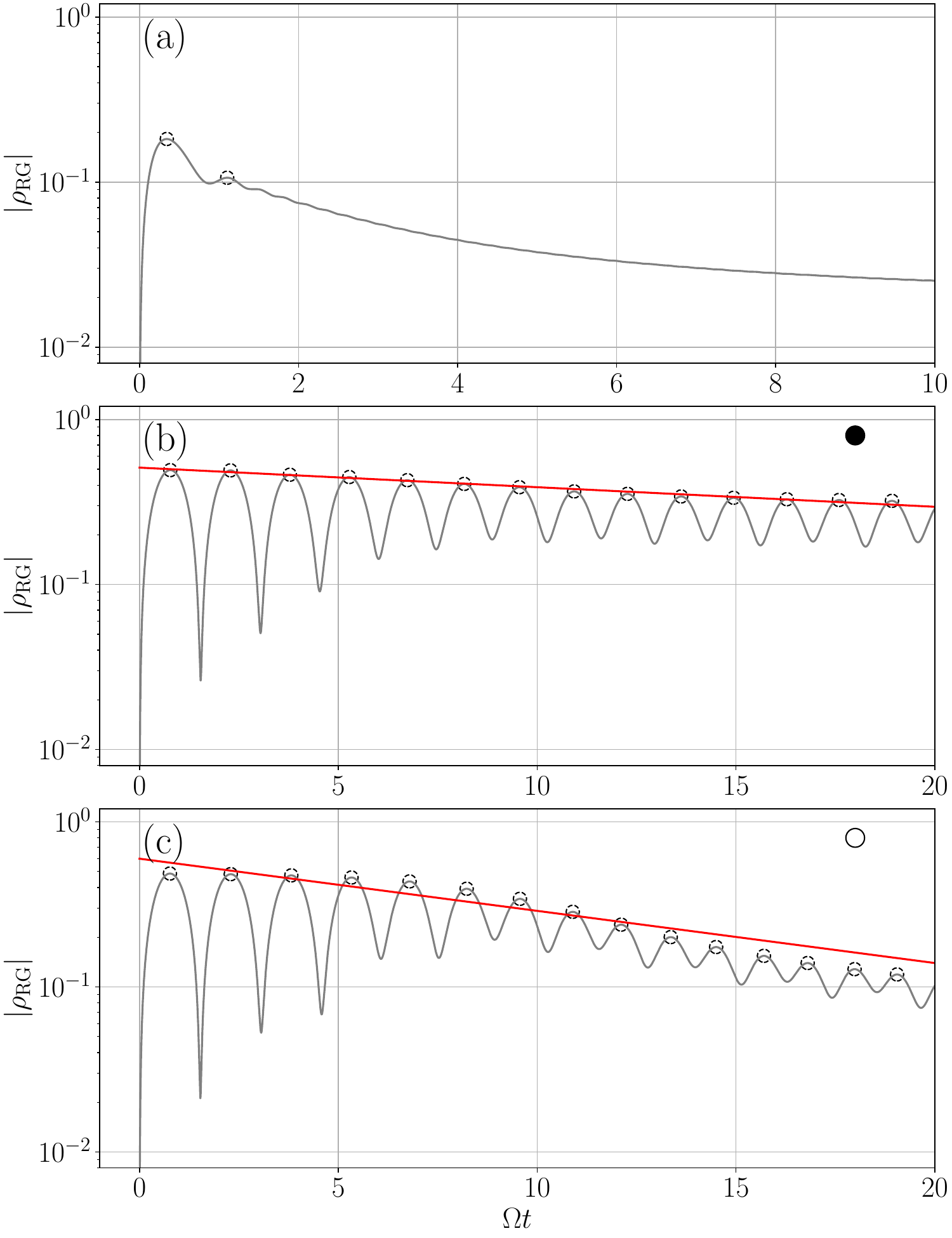}
    \caption{
    Absolute value of $\rho_\mathrm{RG}$ (grey) over time from numerical simulations. The maxima (black circles) are fitted with an exponential decay function (red), yielding the dephasing rate. \brady{Here, the parameters are \eva{(a) $\Delta / \Omega = 28.57$, $\xi / \Omega = 0.07 \cdot10^{-3}$ }(b) $\Delta / \Omega = 1.32$, ${\xi / \Omega = 1.25 \cdot10^{-3}}$ and (c) $\Delta / \Omega = 2.14$, $\xi / \Omega = 0.03 \cdot 10^{-3}$.}
    In (c) the decay does not perfectly follow an exponential form, consequently values of $\gamma_\perp$ extracted from such fits are marked as hollow points in  {later plots when compared to analytic predictions (cf.} Fig.~\ref{fig:dephasing}).
    }
    \label{fig:coherence}
\end{figure}


The algorithm generates the wavefunctions $\psir(x, t)$ and $\psig(x, t)$, where we discretize space in the interval \bradytwo{$x / \rfac \in [0.1, 10.5]$} using ${N=2^{17} \sim 10^5}$ grid points.
Finally, we calculate the coherence with 
\begin{equation}
    \label{eq:RhoRG_dx}
    \rho_\mathrm{RG}(t)=\int\mathrm{d}x\,\psir^*(x,t)\psig(x,t).
\end{equation}

We find the absolute value of the coherence ${|\rho_\mathrm{RG}|}$ to oscillate and decay in amplitude over time, before reaching a steady state. In particular, during this time, we find the maxima of ${|\rho_\mathrm{RG}|}$ to decay exponentially to a reasonable degree of accuracy. We fit this decay with an exponential function of the form $\e^{-\gamma_\perp t}$, where $\gamma_\perp$ is then identified as the dephasing rate.

For typical facilitation parameters, i.e. where ${\Delta / \Omega \gg 1}$, we find this decay to be on the order of, or faster than Rabi oscillations, making a rigorous fitting of the maxima difficult (cf.~Fig.~\ref{fig:coherence}a). For this reason, we investigate the regime where ${\Delta / \Omega \gtrsim 1}$. \bradytwo{Finally, for all simulations we use ${\sigma / x_\mathrm{f} = 0.05}$, which corresponds to typical experimental ratios between tweezer trap spacings and trap widths.}  In Fig.~\ref{fig:coherence}b-c we give two examples of such a simulation, showing Rabi-oscillations damped by an effective dephasing. Note that spontaneous emission was assumed to be negligible on the time scales shown.


\begin{figure}[H]
    \centering
    \includegraphics[width=\columnwidth]{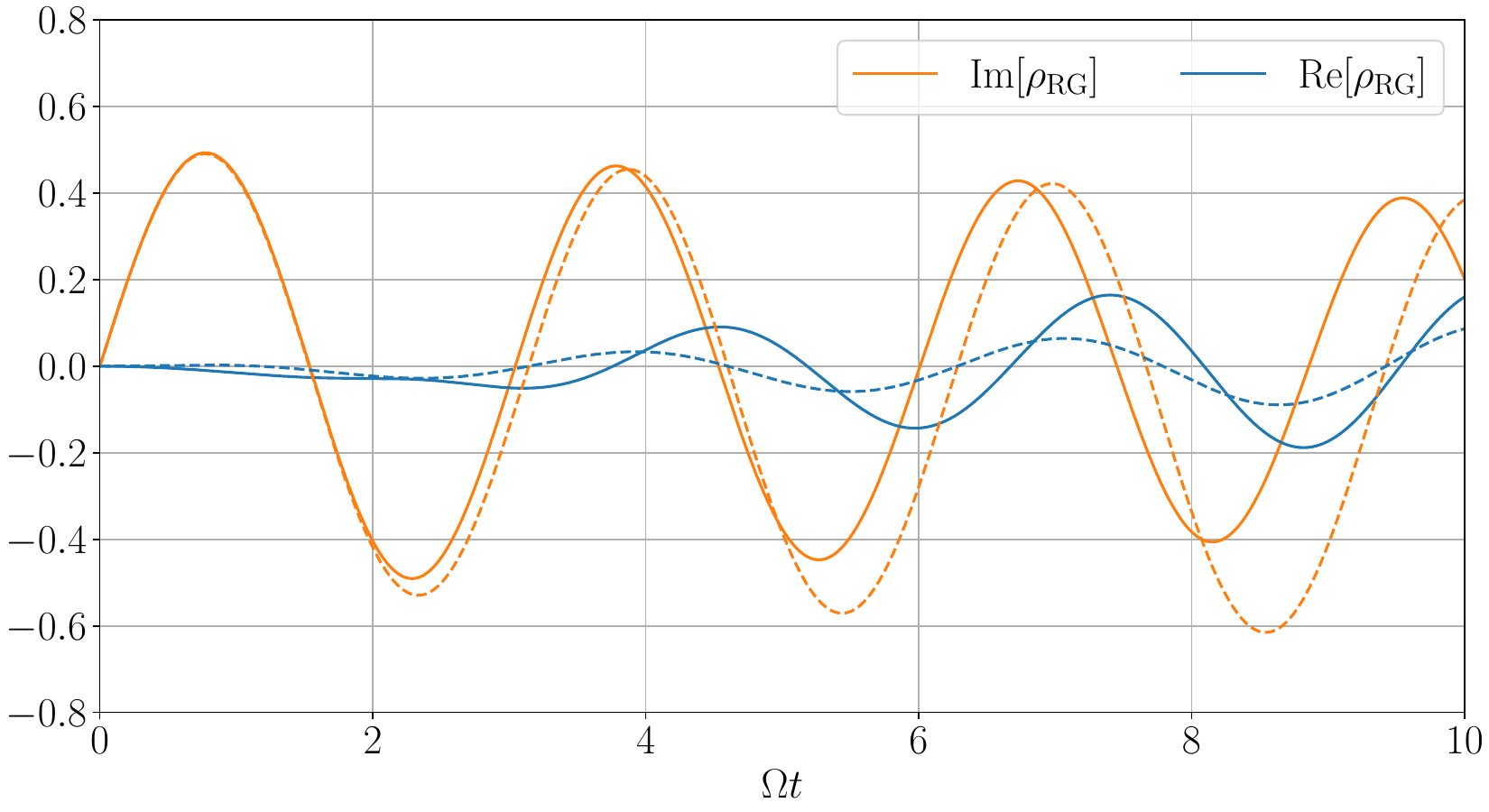}
    \caption{Time resolved imaginary value (orange), and real value (blue) of the coherence $\rho_\mathrm{RG}$ from numerics (solid) and from analytics in first order perturbation theory (cf. eq.~\eqref{eq:Rho_analytic}) (dashed), using \brady{$\Delta / \Omega = 1.32$ and ${\xi / \Omega = 1.25 \cdot10^{-3}}$.}
    }
    \label{fig:coherence-2}
\end{figure}


In Fig.~\ref{fig:coherence-2} we compare the analytic short-time approximations to the full numerics. \mfl{We recognize good agreement between analytics and numerics and see that at the peak values of $\textrm{Im}[\rho_\textrm{RG}](t)$ the real part $\textrm{Re}[\rho_\textrm{RG}](t)$ is indeed small.}

\begin{figure}[H]
    \centering
    \includegraphics[width=\columnwidth]{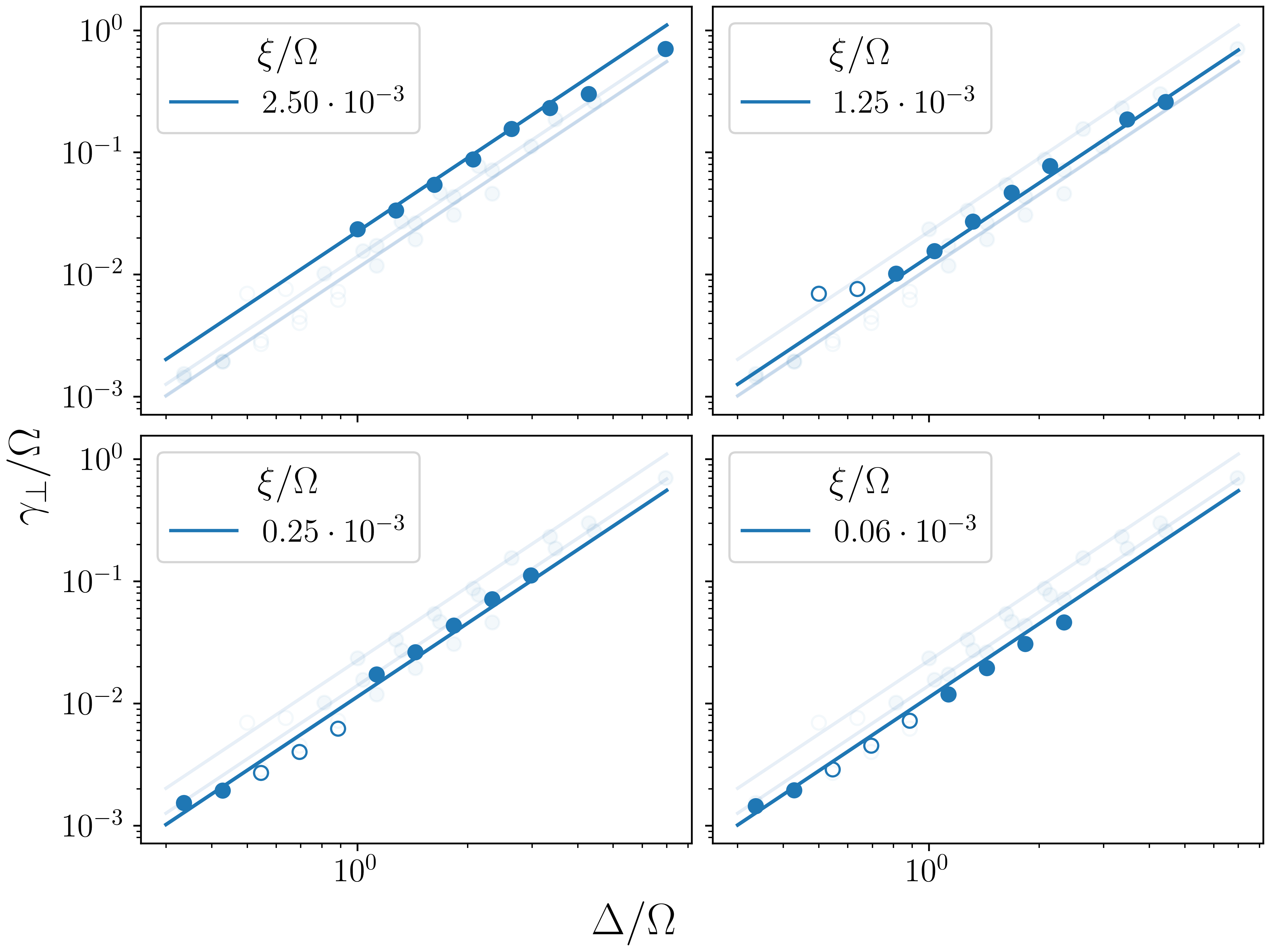}
    \caption{Dephasing rate from exponential fit of numeric simulations. Hollow dots correspond to simulations were $\rho_\mathrm{RG}(t)$ does not perfectly follow an exponential function in time (cf. Fig.~\ref{fig:coherence}c). The lines correspond to eq.~\eqref{eq:DephsingRate}.
    }
    \label{fig:dephasing}
\end{figure}

\mfl{We now compare the analytic predictions for the short-time dephasing rate, eq.~\eqref{eq:DephsingRate}, with numerical simulations. This is shown in Fig.~\ref{fig:dephasing}, where we plotted $\gamma_\perp/\Omega$ as function of $\Delta/\Omega$ for different $\xi/\Omega$. One recognizes very good agreement. 
}
 
\section{Summary}

 In the present paper we  discussed \mfl{the effects of the dipole force acting on the Rydberg-state wavefunction on the  }
 dynamics of a single atom, laser coupled between 
 an initially spatially localized ground state and a high-lying Rydberg state under conditions of Rydberg facilitation, i.e. in the presence of an already excited Rydberg atom at the facilitation distance. In the dipole potential of \mfl{an already excited} Rydberg atom, the atom \mfl{under consideration} experiences a decay of coherences due to dipole forces acting solely on the Rydberg state. 

We model this in second quantization, by  explicitly taking into account the motional degrees of freedom of the atomic ground and Rydberg states, coupled by a laser.  Initially, the atom is assumed in the ground state with a well-localized spatial wavefunction, typical for lattice experiments with Rydberg atoms.
As a result of dipole forces, the Rydberg state wave function is repelled from the external Rydberg atom, and the overlap between ground and Rydberg state wavefunctions decays. Using  a perturbative solution of the coupled equations, we derived an analytic expression for the \mfl{(short time)} dephasing rate $\gamma_\perp$, {which we have benchmarked with numerical simulations for values of $\Delta/\Omega$ up to $ \sim {\cal O}(1)$}.




\subsection*{Authors contribution}
$^*$These authors contributed equally. \daniel{T.S. performed the analytic calculations and E.K. performed the numeric simulations.} D.B. supervised the project with support by M.F. M.F. conceived the project. All authors were involved in the discussion of results and writing of the manuscript.

\subsection*{Acknowledgments}

Financial support from the DFG through SFB TR 185, Project No. 277625399, is gratefully acknowledged. The authors also thank the Allianz f\"ur Hochleistungsrechnen (AHRP) for giving us access to the “Elwetritsch” HPC Cluster.

\bibliographystyle{apsrev4-2}
\bibliography{references}

\end{document}